\documentclass[11pt,onecolumn,journal]{IEEEtran}

\usepackage{epsf,psfrag,amssymb,amsfonts,cite,graphicx,amsmath,color}
\usepackage[mathscr]{eucal}

\def\boxit#1{\vbox{\hrule\hbox{\vrule\kern3pt
        \vbox{\kern3pt#1\kern3pt}\kern3pt\vrule}\hrule}}

\def\reals{ { {\rm  I \kern-0.15em R }  } }
\def\complex{ {\,{{\rm C} \kern-0.50em \raise0.20ex {  |}}\, }}

\def\Rbf{{\bf R}}

\def\Hc{{\cal H}}

\def\Oc{{\cal O}}

\def\Rxx{\Rbf_{\ssstyle X\kern-.1em X}}

\let\ssstyle=\scriptscriptstyle

\def\eg{{\it e.g.,}}

\def\ie{{\it i.e.,\ \/}}
\def\Kout{\setbox1=\hbox{\Huge\bf K}\hbox to
1.05\wd1{\hspace{.05\wd1}
}}
\def\Sout{\setbox1=\hbox{\Huge\bf S}\hbox to 1.05\wd1{\hspace{.05\wd1}
}}

\def\scalefig#1{\epsfxsize #1\textwidth}

\def\ie{{\it i.e.,\ \/}}
\def\eg{{\it e.g.,\ \/}}

\newcommand{\be}{\begin{equation}}
\newcommand{\ee}{\end{equation}}
\newcommand{\bea}{\begin{eqnarray}}
\newcommand{\eea}{\end{eqnarray}}

\def\scalefig#1{\epsfxsize #1\textwidth}
\makeatletter
\def\Lddots{\mathinner{\mkern1mu\raise17\p@\vbox{\kern17\p@\hbox{.}}\mkern2mu
    \raise8\p@\hbox{.}\mkern2mu\raise\p@\hbox{.}\mkern1mu}}
 \makeatother

\outer\def\subsect#1\par{\vskip12pt
plus.07\vsize\penalty-250\vskip0pt plus-.07\vsize
\bigskip\vskip\parskip\message{#1}
\vbox{\smash{\lower9pt\hbox{\kern-8pt\epsfbox{shadedbox.eps}}}}\vskip-\baselineskip
\leftline{\large\bf#1}\nobreak\medskip}

\centerfigcaptionstrue

\begin{document}

\title{\Large\bf
Dynamic Spectrum Access:\\[-0.4em] Signal Processing, Networking, and Regulatory Policy}
\vspace{-8em}
\author{\it \normalsize
Qing Zhao and Brian M. Sadler
\thanks{
Q. Zhao is with the Department of Electrical
and Computer Engineering, University of California, Davis, CA 95616, {\it qzhao@ece.ucdavis.edu}.
B. M. Sadler is with the Army Research Laboratory, Adelphi, MD 20783, {\it bsadler@arl.army.mil}.
This work was supported by the
Army Research Laboratory CTA on Communication and Networks under Grant
DAAD19-01-2-0011.
}
}

\markboth{Submitted to {\it IEEE Signal Processing Magazine}, September, 2006}{Zhao and Sadler}
\maketitle

\vspace{-4em}




\thispagestyle{empty}

\begin{center}
\begin{minipage}{6in}
\setlength{\baselineskip}{8pt}
\setlength{\parskip}{0pt}
{\footnotesize
\hrulefill\\[-1em]
\vspace{-1em}
\tableofcontents
}
\hrulefill
\end{minipage}
\end{center}

\newpage

\setcounter{page}{1}

\section{Introduction}

\subsection{Spectrum Reform: Concepts and Taxonomies}

\begin{figure}[htb]
\centerline{
\begin{psfrags}
\scalefig{0.5}\epsfbox{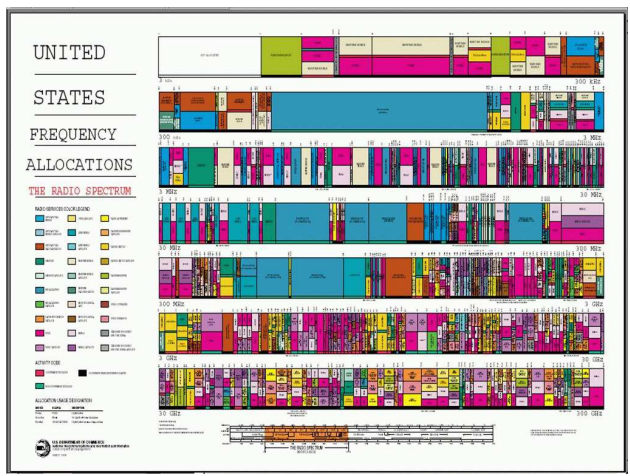}
\end{psfrags}}
\caption{Radio frequency allocation chart in US.}
\label{fig:chart}
\end{figure}

There is a common belief
that we are running out of usable radio frequencies.
The overly crowded US frequency allocation chart (see Figure~\ref{fig:chart}) and the multi-billion-dollar price for a 20MHz frequency band at
the European 3G spectrum auction have certainly strengthened this belief.

Are we truly approaching the capacity of
the radio spectrum?   Actual spectrum usage measurements obtained
by the FCC's Spectrum Policy Task Force \cite{FCC:02TaskForceRpt} tell a different story: at any given
time and location, much of the prized spectrum lies idle. This paradox indicates that spectrum shortage
results from the spectrum management policy rather than the physical scarcity of usable frequencies.
Analogous to idle slots in a static TDMA system with bursty traffic, idle frequency bands are inevitable under
the current static spectrum allotment policy that grants exclusive use to licensees.

The underutilization of spectrum has stimulated a flurry of exciting activities in engineering, economics,
and regulation communities in searching for better spectrum management policies. The diversity of the envisioned
spectrum reform ideas is manifested in the number of technical terms coined so far: dynamic spectrum access vs.
dynamic spectrum allocation, spectrum property rights vs. spectrum commons, opportunistic spectrum access vs.
spectrum pooling, spectrum underlay vs. spectrum overlay. Compounding the confusion is the use of cognitive
radio as a synonym of dynamic spectrum access.  As an initial
attempt at unifying the terminology, we provide below a taxonomy.

\begin{figure}[htb]
\centerline{
\begin{psfrags}
\scalefig{1}\epsfbox{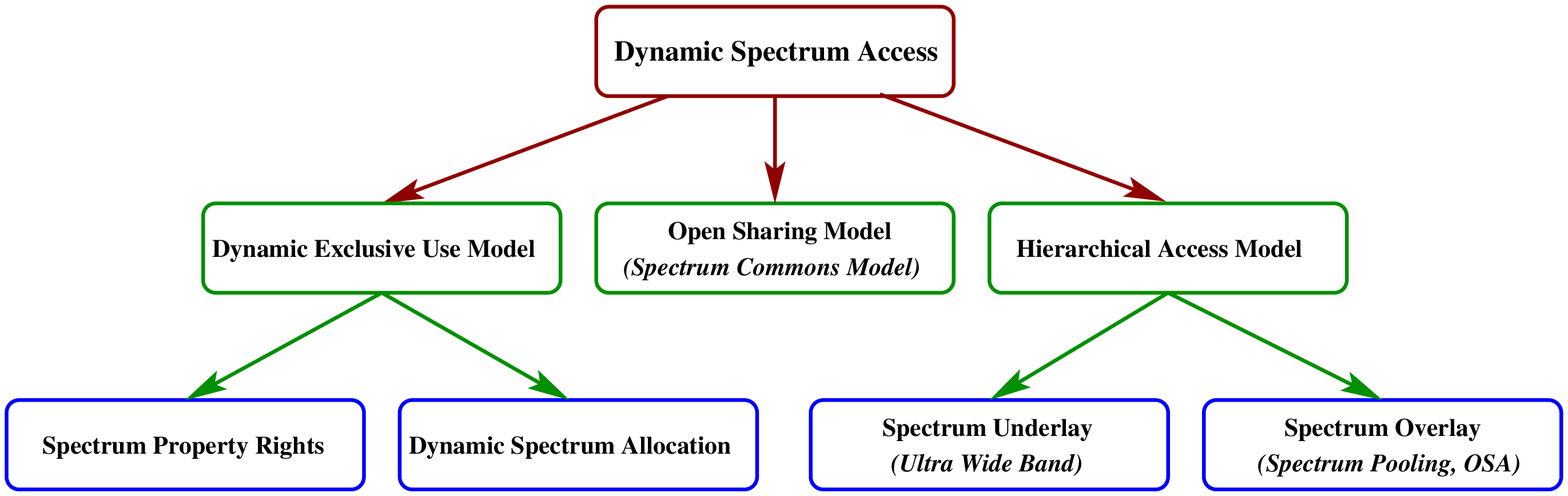}
\end{psfrags}
}
\caption{A taxonomy of dynamic spectrum access}
\label{fig:terms}
\end{figure}

\noindent{\bf Dynamic Spectrum Access}~~~
Standing for the opposite of the current static spectrum management policy, the term ``dynamic spectrum access'' has
broad connotations that encompass various approaches to spectrum reform. The diverse ideas presented at the
first {\it IEEE Symposium on New Frontiers in Dynamic Spectrum Access Networks (DySPAN)} \cite{DySPAN05} suggest the extent of this term.
As illustrated in Figure~\ref{fig:terms}, dynamic spectrum access strategies can be generally categorized under three models.
\begin{enumerate}
\item {\it Dynamic Exclusive Use Model}\\
This model maintains the basic structure of the current spectrum regulation policy: spectrum bands are licensed
to services for exclusive use. The main idea is to introduce flexibility
to improve spectrum efficiency. Two approaches have been proposed under this model:
{\it spectrum property rights}~\cite{Coase:59,Hatfield&Weiser:05DySPAN}
and {\it dynamic spectrum allocation} \cite{Xu&etal:00LCN}. The former approach allows
licensees to sell and trade spectrum and to freely choose technology. Economy and market will thus play a more important role
in driving toward the most profitable use of this limited resource. Note that even though licensees have the right to lease or
share the spectrum for profit, such sharing is not mandated by the regulation policy.

The latter approach, dynamic spectrum allocation, was brought forth by the
European DRiVE project~\cite{Xu&etal:00LCN}. It aims to improve spectrum efficiency through dynamic spectrum assignment
by exploiting the spatial and temporal traffic statistics of different services. In other words, in a given region and
at a given time, spectrum is allocated to services according to a chart similar to that in Figure~\ref{fig:chart}.
This allocation, however, varies at a much faster scale than the current policy.

Based on an exclusive-use model, however, these approaches cannot eliminate
white space in spectrum resulting from the bursty nature of wireless traffic.

\item {\it Open Sharing Model}\\
Also referred to as spectrum commons \cite{Benkler:97,Lehr&Crowcroft:05DySPAN}, this model employs
open sharing among peer users as the basis for managing a spectral region.
Advocates of this model draw
support from the phenomenal success of wireless services operating in the unlicensed ISM band (\eg WiFi).
Centralized \cite{Raman&etal:05DySPAN,Ileri&etal:05DySPAN} and distributed \cite{Chung&etal:03ISIT,Etkin&etal:05DySPAN,Huang&etal:05DySPAN}
spectrum sharing strategies have been initially investigated to address technological challenges under this spectrum management model.

\item {\it Hierarchical Access Model}\\
Built upon a hierarchical access structure with primary and secondary users, this model
can be considered as a hybrid of the above two. The basic idea is to open licensed spectrum to
secondary users and limit the interference perceived by primary users (licensees). Two approaches to spectrum sharing between
primary and secondary users have been considered: {\it spectrum underlay} and {\it spectrum overlay}.

The underlay approach imposes severe constraints on the transmission power of secondary users
so that they operate below the noise floor of primary users. By spreading transmitted signals
over a wide frequency band (UWB), secondary users can potentially achieve short-range high data rate with extremely low
transmission power. Based on a worst-case assumption that primary users transmit all the time,
this approach does not exploit spectrum white space.

Spectrum overlay was first envisioned by Mitola \cite{Mitola:99} under the term
``spectrum pooling'' and then investigated by
the DARPA XG program \cite{DARPA:XG} under the term ``opportunistic spectrum access (OSA)''.
Differing from spectrum underlay, this approach does not necessarily impose severe restrictions on the transmission power
of secondary users, but rather on when and where they may transmit.
It directly targets at spatial and temporal spectrum white space by allowing secondary users
to identify and exploit local and instantaneous spectrum availability in a nonintrusive manner.

Compared to the dynamic exclusive use and open sharing models, this hierarchical model is perhaps the most compatible
with the current spectrum management policy and legacy wireless systems. Furthermore, the underlay and overlay
approaches can be employed simultaneously to further improve spectrum efficiency.

\end{enumerate}

\noindent{\bf Cognitive Radio}~~~
Software-Defined Radio and cognitive radio were coined by Mitola in 1991 and 1998, respectively.
Software-defined radio, sometimes shortened to software radio,
is generally a multi-band radio that supports multiple air interfaces and protocols and is reconfigurable through software run on
DSP or general-purpose microprocessers \cite{Mitola:book}. Cognitive radio, built upon a software radio platform, is a context-aware
intelligent radio capable of autonomous reconfiguration by learning and adapting to the communication environment \cite{Mitola:98}.
While dynamic spectrum access is certainly an important application of cognitive radio, cognitive radio represents a much broader
paradigm where many aspects of communication systems can be improved via cognition.

\subsection{Opportunistic Spectrum Access: Basic Components}

In this article, we focus on the overlay approach under the hierarchical access model (see Figure~\ref{fig:terms}). The term
of Opportunistic Spectrum Access (OSA) will be adopted throughout.

Basic components of OSA include spectrum opportunity identification, spectrum opportunity exploitation, and
regulatory policy. The opportunity identification module is responsible for
accurately identifying and intelligently tracking idle frequency bands that are dynamic in both time and space.
The opportunity exploitation module takes input from the opportunity identification module and decides
whether and how a transmission should take place. The regulatory policy defines the basic etiquettes for secondary users
to ensure compatibility with legacy systems.

The overall design objective of OSA is to provide sufficient benefit to secondary users while protecting spectrum licensees
from interference. The tension between the secondary users' desire for performance and the primary users' need for protection
dictates the interaction across opportunity
identification, opportunity exploitation, and regulatory policy. The optimal design of OSA
thus calls for a cross-layer approach that integrates signal processing and networking with regulatory policy making.

In this article, we aim to provide an overview of challenges and
recent developments in both technological and regulatory aspects of OSA.
The three basic components of OSA will be discussed in Sections~\ref{sec:identification} to \ref{sec:policy},
respectively.

\subsection{An Example of OSA Networks}
\label{sec:example}

To illustrate the basic technical issues in OSA, we often resort to the following example of OSA networks.
The design challenges, tradeoffs, and many existing results presented in this article, however, apply to general
OSA networks.

We consider a spectrum
consisting of $N$ channels. Here we use the term ``channel'' broadly.
A channel can be a frequency band with certain bandwidth,  a collection of spreading codes in a CDMA network,
or a set of tones in an OFDM system.
We assume that interference across channels is negligible. Thus, a secondary
user transmitting over an available channel does not interfere with primary users using other channels.
This assumption imposes constraints on the modulation of secondary users, as will be discussed in
Section~\ref{sec:OFDM}.

These $N$ channels are allocated to a network of primary users who communicate according to a synchronous slot structure.
The traffic statistics of the primary network are such that
the occupancy of these $N$ channels follows a Markov process with $2^N$
states, where the state is defined as the availability (idle or busy) of each channel.
Overlayed with this primary network is an ad hoc secondary network where users
seek spectrum opportunities in these $N$ channels independently.
In each slot, a secondary user may choose a set of channels to sense
and a set of channels to access based on the sensing outcome.
Accessing an idle channel leads to bit delivery, and accessing a busy channel results in a collision
with primary users.
The objective is to maximize the throughput of secondary users while
limiting the probability of colliding with primary users below a prescribed
level.
To simply notation and illustrate the basic idea, we assume that
a secondary use can choose only one channel to sense and possibly access in each slot.
Extensions to general cases are often straightforward.

\section{Spectrum Opportunity Identification}
\label{sec:identification}

Spectrum opportunity identification is crucial to OSA in order to achieve nonintrusive communication.
In this section, we identify basic functions of the opportunity identification module.

\subsection{Spectrum Opportunity and Interference Constraint: Definitions and Implications}

\subsubsection{Spectrum Opportunity}

Before discussing spectrum opportunity identification, a rigorous definition of spectrum opportunity is necessary.
Intuitively, a channel can be considered as an opportunity if it is not currently used by primary users.
In a network with geographically distributed
primary transmitters and receivers, however, the concept of spectrum opportunity is more
involved than it at first may appear \cite{Zhao:07ICASSP}.

\begin{figure}[htb]
\centerline{
\begin{psfrags}
\psfrag{A}[c]{ $A$}
\psfrag{B}[c]{ $B~$}
\psfrag{i}[c]{ ~~~~~~~~~~~Interference}
\psfrag{ra}[r]{\Large {$r_{tx}$}}
\psfrag{rb}[c]{\Large {$~~r_{rx}$}}
\psfrag{tx}[l]{ Primary Tx}
\psfrag{rx}[l]{ Primary Rx}
\scalefig{0.5}\epsfbox{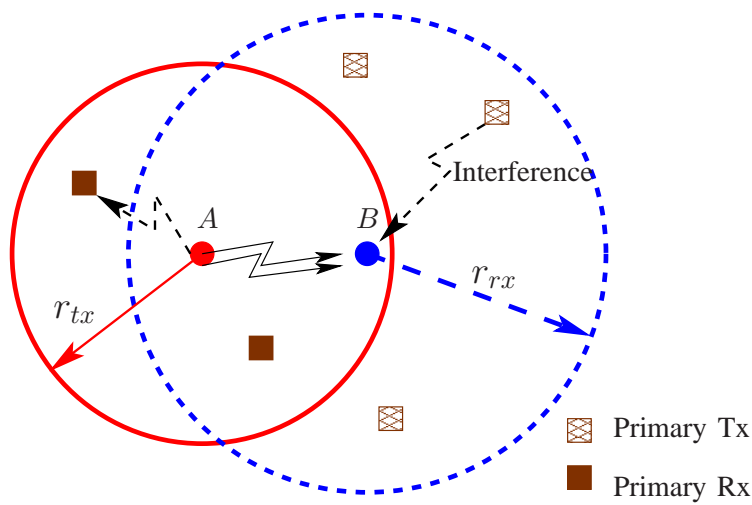}
\end{psfrags}}
\caption{Definition of spectrum opportunity.}
\label{fig:DefSO}
\end{figure}

With the help of Figure~\ref{fig:DefSO}, we identify conditions for a channel to be
considered as an opportunity. Consider a pair of secondary users where $A$ is the transmitter and $B$ its
intended receiver. A channel is an opportunity to $A$ and $B$ if they can communicate successfully
over this channel while limiting the interference to primary users below a prescribed level determined by
the regulatory policy. This means that receiver $B$ will {\it not be affected} by primary transmitters and transmitter $A$ will
{\it not interfere with} primary receivers.

To illustrate the above conditions, we consider monotonic and uniform signal attenuation and omnidirectional antenna.
In this case, a channel is an opportunity to $A$ and $B$ if
no primary users within a distance of $r_{tx}$ from $A$ are receiving and
no primary users within a distance of $r_{rx}$ from $B$ are transmitting
over this channel (see Figure~\ref{fig:DefSO}). Clearly,
$r_{tx}$ is determined by the secondary users' transmission power and the maximum allowable interference to primary users
while $r_{rx}$ is determined by the primary users' transmission power and the secondary users' interference tolerance.
They are generally different.

We make the following remarks regarding the above definition of spectrum opportunity.
\begin{itemize}
\item Spectrum opportunity is a local concept defined with respect to a particular pair of secondary users. It depends on
the location of not only the secondary transmitter but also the secondary receiver. For multicast and broadcast, spectrum
opportunity is open for interpretation.
\item Spectrum opportunity is determined by the communication activities of primary users rather than that of secondary users.
Failed communications
caused by collisions among secondary users do not disqualify a channel from being an opportunity.
\end{itemize}

\subsubsection{Interference Constraint}

How to impose interference constraints is a complex regulatory issue. Restrictive constraints
may marginalize the potential gain of OSA while loose constraints may affect the compatibility with
legacy systems.

Generally speaking, an interference constraint should implicitly or explicitly specify at least two parameters: the maximum
interference power level $\eta$ perceived by an active primary receiver and the maximum probability
$\zeta$ that the interference level at an active primary receiver may exceed $\eta$ \cite{Zhao:07ICASSP}.
The first parameter, $\eta$, can be considered as specifying the noise floor of primary users; interference
below $\eta$ does not affect primary users while interference above $\eta$ results in a collision. It is
thus inherent to the definition of spectrum opportunity and determines the transmission power of secondary
users as discussed in Section~\ref{sec:OFDM}.

The second parameter, $\zeta$, specifies the maximum collision probability allowed by primary users. Given
that errors in spectrum opportunity detection are inevitable, a positive value of $\zeta$ is necessary
for secondary users to ever be able to exploit an opportunity. As discussed in Section~\ref{sec:access},
$\zeta$ determines a secondary transmitter's access decision based on imperfect
spectrum opportunity detection. A cautionary aspect of collision probability is the probability space
over which it is defined. As shown in \cite{Zhao:07ICASSP}, collision probabilities defined with respect
to different probability space have different implications and offer different levels of protection to
primary users.

While an interference constraint specified by $\{\eta,\zeta\}$ should be imposed on the aggregated transmission
activities of all secondary users, each secondary user needs to know the node-level constraint in order to
choose transmission power and make access decisions. The translation from a network-level interference constraint
to a node-level one depends on the geolocation and traffic of secondary users.

\subsection{Spectrum Opportunity Detection}
\label{sec:detection}

\subsubsection{Signal Processing and Networking Techniques for Opportunity Detection}

From the definition of spectrum opportunity illustrated in Figure~\ref{fig:DefSO}, it is clear that in a general network setting,
spectrum opportunity detection needs to be performed jointly by the secondary transmitter and receiver.
It thus has both signal processing and networking aspects.

Consider the OSA network example given in Section~\ref{sec:example}.
At the beginning of each slot, a pair of communicating secondary users need to determine whether a chosen channel is an opportunity
in this slot. Ignore for now the contention among secondary users. One approach to opportunity detection is as follows
\cite{Zhao&etal:05Asilomar,Zhao&etal:07JSAC}.
The transmitter
first detects the receiving activities of primary users in its neighborhood (see Figure~\ref{fig:DefSO}). If the channel is available
(no primary receivers nearby),
it transmits a short request-to-send (RTS) message to the receiver. The receiver, upon successfully receiving the RTS,
knows that the channel is also available at the receiver side (no primary transmitters nearby
since RTS has been successfully received) and replies with a clear-to-send (CTS) message.
A successful exchange of RTS-CTS completes
opportunity detection and is followed by data transmission. As detailed in \cite{Zhao&etal:07JSAC}, when RTS and CTS are transmitted
using carrier sensing, this RTS-CTS exchange has dual functions. Besides facilitating opportunity detection, it also addresses contention
among secondary users and mitigates the
hidden and exposed terminal problem as in a conventional communication network \cite{Tanenbaum:96book}.

What remains to be solved is the detection of the receiving activities of primary users by the secondary transmitter.
Without assuming cooperation from primary users, primary receivers are much harder to detect than primary transmitters.
For the application of secondary wireless services operating in the TV bands, Wild and Ramchandran \cite{Wild&Ramchandran:05DySPAN}
proposed to exploit the local oscillator leakage power emitted by the RF front end of TV receivers to detect the presence of
primary receivers. The difficulty of this approach lies in its short detection range and long detection time to achieve accuracy.
It is proposed in \cite{Wild&Ramchandran:05DySPAN} that low-cost sensors be deployed close to primary receivers for spectrum opportunity
detection.

\begin{figure}[htb]
\centerline{
\begin{psfrags}
\psfrag{A}[c]{ $A$}
\psfrag{X}[c]{ $X$}
\psfrag{Y}[c]{ $Y$}
\psfrag{tx}[l]{ Primary Tx}
\psfrag{rx}[l]{ Primary Rx}
\psfrag{ra}[r]{\Large {$r_{tx}$}}
\psfrag{R}[l]{\Large {$R_p+r_{tx}$}}
\scalefig{0.4}\epsfbox{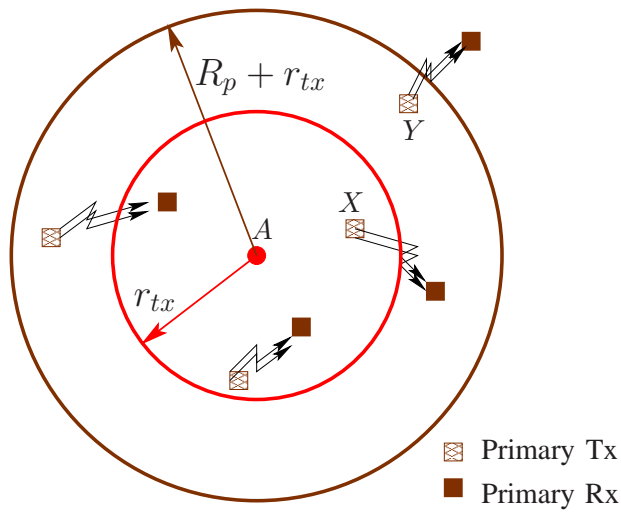}
\end{psfrags}}
\caption{Spectrum opportunity detection.}
\label{fig:OD}
\end{figure}

Another approach is to transform the problem of detecting primary receivers to detecting primary transmitters. Let $R_p$ denote the
transmission range of primary users, \ie primary receivers are within $R_p$ distance to their transmitters. A secondary transmitter
can thus determine that a channel is available if no primary transmitters are detected within a distance of $R_p+r_{tx}$ as illustrated
in Figure~\ref{fig:OD}. This approach, however, may lead to overlooked opportunities.
As shown in Figure~\ref{fig:OD}, the transmission activities of
primary nodes $X$ and $Y$ may prevent $A$ from accessing an opportunity even though the intended receivers
of $X$ and $Y$ are outside the interfering range $r_{tx}$ of $A$.

Even though this approach is conservative,
it reduces spectrum opportunity detection to a classic signal processing problem. As discussed in
\cite{Cabric&etal:04Asilomar}, based on the secondary user's knowledge of the signal characteristics
of primary users, three traditional signal detection techniques can be employed: matched filter, energy
detector (radiometer), and cyclostationary feature detector. A matched filter performs coherent detection. It requires
only $\Oc(1/SNR)$ samples to achieve a given detection power but relies on synchronization and {\it a priori}
knowledge of primary users' signaling (the detector might also use known flags or training symbols in the primary users' signal).
On the other hand, the non-coherent energy detector requires only basic
information of primary users' signal characteristics but suffers from long detection time: $\Oc(1/SNR^2)$
samples are needed for a given detection power. Cyclostationary feature detector
can improve the performance over an energy detector by exploiting an inherent periodicity in the primary users' signal.
Details of this detector can be found in \cite{Gardner:88COM}.

While classic signal detection techniques exist in the literature, detecting primary transmitters in a dynamic
wireless environment with noise uncertainty, shadowing, and fading is a challenging problem as articulated
in \cite{Sahai&etal:04Allerton}. To improve detection accuracy, researchers have suggested cooperative
spectrum sensing \cite{Cabric&etal:04Asilomar,Ghasemi&Sousa:05DySPAN,Mishra&etal:06ICC}. The basic idea is to
overcome shadowing and multipath fading by allowing neighboring secondary users to exchange sensing information
through a dedicated control channel. The overhead associated with sensing information exchange, the feasibility
of a control channel, and the applicability to OSA networks with fast varying spectrum usage require further
investigation.

\subsubsection{Performance Characteristics of Spectrum Opportunity Detector}

The spectrum opportunity detector of secondary users determines whether a chosen channel
is an opportunity. It can be considered as performing a binary hypotheses
test where the null hypothesis
$\Hc_0$ indicates an opportunity and hypothesis $\Hc_1$ is the alternative.
If the detector mistakes $\Hc_0$ for $\Hc_1$, a false alarm occurs, and
a spectrum opportunity is overlooked by the detector. On the other hand, when the detector mistakes $\Hc_1$ for $\Hc_0$,
we have a miss detection.
Let $\epsilon$ and $\delta$ denote, respectively, the probabilities of false alarm
and miss detection.
The performance of the detector is
specified by the Receiver Operating Characteristics (ROC) curve
which gives $1-\delta$ (probability of detection or detection power)
as a function of $\epsilon$.
As illustrated in Figure~\ref{fig:access}, a smaller false alarm probability $\epsilon$ implies a larger
miss detection probability $\delta$.

The ROC curve of the detector is the interface between the opportunity identification module and the
opportunity exploitation module.
As discussed in Section~\ref{sec:access}, access decisions of secondary users should take into account the
operating characteristics of the opportunity detector. Different operating points $\delta\in(0,1)$ on the ROC curve lead to
different optimal access strategies and different throughput performance of secondary users. A joint design
of the opportunity identification module and the opportunity exploitation module is thus necessary to achieve
the optimal performance \cite{Chen&etal:06Asilomar}.

\subsection{Spectrum Opportunity Tracking}

Due to hardware limitation and energy cost associated with spectrum monitoring, a secondary user may not be able to
sense all channels in the spectrum simultaneously. In this case,
the secondary user needs a sensing strategy for intelligent channel selection to track the rapidly varying spectrum
opportunities.
The purpose of the sensing strategy is twofold: catch a spectrum opportunity for immediate access and
obtain statistical information on spectrum occupancy so that more rewarding sensing decisions can be made in the future.
A tradeoff has to be reached between these two often conflicting objectives.

Consider again the OSA network example in Section~\ref{sec:example}. A simple static sensing strategy
would choose the channel most likely to be available (weighted by its bandwidth) based on
the stationary distribution of the underlying Markov process.
In this case, the secondary user simply
waits on a particular channel predetermined by the spectrum occupancy statistics and the channel bandwidths.
Missing in
this approach is that every sensing outcome provides information on the state of the underlying Markov process. Channel selection
should be based on the {\it conditional} distribution of channel availability that exploits the whole history of sensing outcomes.

\begin{figure}[htb]
\centerline{
\begin{psfrags}
\scalefig{0.45}\epsfbox{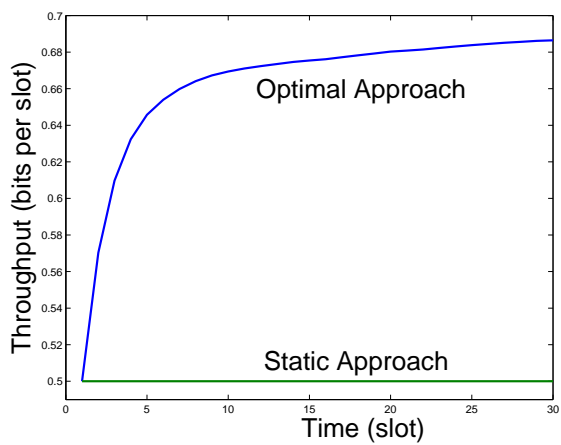}
\end{psfrags}
}
\caption{Spectrum opportunity tracking: a sequential decision-making problem.}
\label{fig:sensing}
\end{figure}

The optimal sensing strategy is thus one of sequential decision-making that achieves the best
trade-off between gaining immediate access
in the current slot and gaining system state information for
future use. In \cite{Zhao&etal:05DySPAN,Zhao&etal:07JSAC}, the optimal sensing strategy has been formulated and
addressed within the framework of Partially Observable Markov Decision Process (POMDP). Based on the results
in \cite{Zhao&etal:05DySPAN,Zhao&etal:07JSAC}, we illustrate
the potential
gain of optimally using the observation history with a simple numerical example where we have three
channels. Plotted in Figure~\ref{fig:sensing} is the throughput of the secondary user as a function of time.
We see from this figure that the performance of the optimal approach based on the POMDP framework
improves over time, which results from the increasingly
accurate information
on the system state drawn from accumulating observations. Approximately $40\%$ improvement is achieved over the static approach.

It is therefore apparent that a simple yet sufficiently accurate statistic model of spectrum occupancy is crucial to the
efficiency of spectrum opportunity tracking. Spectrum monitoring test-beds \cite{Geirhofer&etal:06MILCOM} and cognitive radio
prototypes \cite{Kleider&etal:05MILCOM} are being developed
by researchers from both academe and industry. They provide empirical data for the statistical modeling of spectrum occupancy.
Results in \cite{Geirhofer&etal:06MILCOM,Geirhofer&etal:06TAPAS} demonstrate that 802.11b has heavy-tailed idle periods, and a continuous-time
semi-Markov process model is proposed.

\section{Spectrum Opportunity Exploitation}
\label{sec:exploitation}

After spectrum opportunities are detected, secondary users need to decide whether and how to exploit them.
Specific issues include whether to transmit given that opportunity detectors may make mistakes, what modulation and transmission
power to use, and how to share opportunities among secondary users to achieve a network-level objective.

\subsection{Whether to Access}
\label{sec:access}

A secondary user needs an access strategy to determine whether to transmit over a particular channel
based on the detection outcome.
Had the spectrum detector been perfect, the design of the access strategy would have been straightforward.
In the presence of detection errors, the access strategy is complicated by the need to decide how much and when to trust the detector.
The trade-off is between minimizing overlooked spectrum opportunities and
avoiding collisions with primary users,
and a good access strategy should balance the desire to aggressively pursue opportunities with the need to avoid
excessive collisions with primary users.

The optimal access strategy should take into account the operating characteristics of the spectrum detector.
Intuitively, when the miss detection probability of the detector is large (\ie a busy channel is often detected as an opportunity),
the access policy
should be conservative to avoid excessive collisions. On the other hand, when the detector has a high false alarm probability, the
access policy should be aggressive to reduce overlooked spectrum opportunities. For any given operating point $\delta$ on the ROC curve,
exactly how aggressive or how conservative the optimal access policy should be is, however, not a trivial problem.

\begin{figure}[htb]
\centerline{
\begin{psfrags}
\psfrag{e}[c]{ $\epsilon$}
\psfrag{d}[c]{ $~~~~1-\delta$}
\psfrag{z}[r]{ $1-\zeta$}
\psfrag{o}[c]{ $\delta>\zeta$}
\psfrag{a}[c]{ $\delta<\zeta$}
\psfrag{C}[c]{\small conservative}
\psfrag{A}[c]{\small aggressive}
\psfrag{O}[c]{\small ~~~~~~~~~~~~~~optimal ($\delta^*=\zeta$)}
\scalefig{0.35}\epsfbox{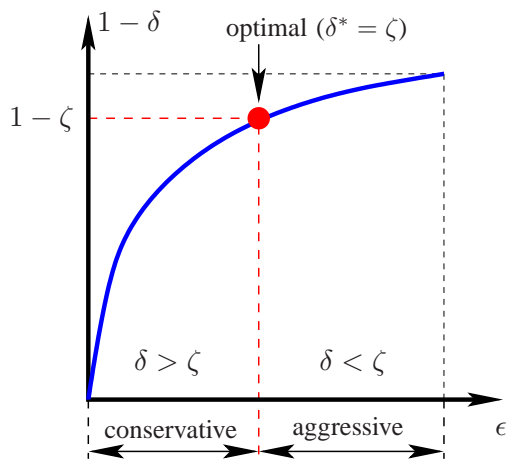}
\end{psfrags}}
\caption{Impact of detector operating point on the optimal access strategy.}
\label{fig:access}
\end{figure}

For the OSA network given in Section~\ref{sec:example},
a separation principle based on a POMDP framework has been established in \cite{Chen&etal:06Asilomar}
that leads to a closed-form characterization of the optimal
access strategy jointly designed with the sensing strategy for any operating point $\delta\in(0,1)$
of the spectrum detector.
As illustrated in Figure~\ref{fig:access}, the ROC curve of the detector is
partitioned into two regions by
the maximum allowable collision probability $\zeta$.
When the detector operates at $\delta>\zeta$, there is a high chance that a busy channel is detected as an opportunity.
If the access strategy completely trusts the detector, a collision occurs if and only if the detector fails to detect the
presence of primary receivers (a miss detection occurs). The collision probability is then given by the miss detection probability $\delta$,
which is larger than $\zeta$. This suggests that when the detector operates in this region, the optimal
access policy should be conservative. Indeed, as shown in \cite{Chen&etal:06Asilomar},
when the channel is detected to be busy, the user should always refrain from transmission;
even when the channel is detected to be available, it should only transmit with probability $\frac{\zeta}{\delta}<1$.

On the other hand, in the region
of $\delta<\zeta$, false alarms are more likely to happen. The user should adopt an aggressive access strategy:
when the channel is detected to be available, always transmit; even when the channel is detected to be busy,
one should still transmit with probability $\frac{\zeta-\delta}{1-\delta}>0$.

Interestingly, the optimal joint design of detector operating point and sensing and access strategies
requires that the detector operates at the transition point $\delta^*=\zeta$ and the optimal access strategy is to simply trust the detector:
access if and only if the channel is detected to be available.
In other words, the access strategy does not need to be conservative nor aggressive to balance the occurrence of false alarms and
miss detections.

\subsection{How to Access}
\label{sec:OFDM}

Modulation and power control in OSA networks also present unique challenges not encountered in the conventional
wired or wireless networks. Since secondary users often need to transmit over noncontiguous frequency bands, orthogonal
frequency division multiplexing (OFDM) has been considered as an attractive candidate for modulation in OSA networks
\cite{Weiss&Jondral:04COMM,Berthold&Jondral:05DySPAN,Tang:05DySPAN}.
The reconfigurable subcarrier structure of OFDM allows secondary users to efficiently fill the spectral gaps left by
primary users without causing unacceptable interference.
The FFT component of OFDM can also be used by the energy detector of secondary users for opportunity
detection.

There are, however, several constraints in designing an OFDM overlay system. First, the subcarrier spacing and
symbol interval need to match with the spectral and temporal duration of spectrum opportunities
\cite{Berthold&Jondral:05DySPAN}.
Second, spectrum leakage caused by signal truncation in the time domain and nonlinearity of the transmitter's power
amplifier needs to be controlled to ensure nonintrusive communication. Carefully designed pulse shaping can reduce
the smearing effect in the frequency domain induced by time-domain truncation. Subcarriers adjacent to channels occupied
by primary users may be nullified or allocated with low power to meet the interference requirement, giving power allocation
an interesting twist. Furthermore, the impact of nulled subcarriers on the peak-to-average-power ratio of the transmitted
OFDM signal requires careful study.

Transmission power control is another complex issue in OSA networks.
To illustrate the basic parameters that affect power control, we ignore shadowing and fading and
focus on a single secondary user.
Consider first that the secondary transmitter $A$ is able to detect the presence of primary receivers within a distance
of $d$ (see Figure~\ref{fig:DefSO} with $r_{tx}$ replaced by $d$).
The transmission power $P_{tx}$ of $A$ should ensure that the signal strength at $d$
away from $A$ is below the maximum allowable interference level $\eta$. This leads to
\[
P_{tx}\le \eta d^\alpha,
\]
where $\alpha$ is the path attenuation factor. The above equation indicates how the maximum transmission power of a secondary user
depends on the detection range $d$ of its spectrum detector, the prescribed maximum interference level $\eta$,
and the path loss factor $\alpha$.

When the secondary user can only detect the presence
of primary transmitters within a distance of $d$ (see Figure~\ref{fig:OD} with $R_p+r_{tx}=d$), we have,
\[
P_{tx}\le \eta (d-R_p)^\alpha,
\]
where $R_p$ is the transmission range of primary users. In other words, power control for secondary users should also
take into account the transmission power of primary users. When we consider shadowing, fading, and interference aggregation
due to simultaneous transmissions from multiple secondary users, a probabilistic model may be necessary to address power control in OSA
networks \cite{Zhao:07ICASSP}.

\subsection{Opportunity Sharing among Secondary Users}

So far we have been focusing on individual noncooperative secondary users.
In the context of exploiting locally unused TV broadcast bands, spatial spectrum opportunity sharing among
secondary users has been investigated (see
\cite{Zheng&Peng:05ICC,Wang&Liu:05VTC,Sankaranarayanan&etal:05DySPAN,Steenstrup:05DySPAN} and references therein).
For this type of application,
spectrum opportunities are considered static or slowly varying in time. Realtime opportunity identification is not as critical
a component as in applications that exploit temporal spectrum opportunities (as in the example
given in Section~\ref{sec:example}).
It is often assumed that spectrum opportunities at any location over the entire spectrum are known.

\begin{figure}[htb]
\centerline{
\begin{psfrags}
\psfrag{C1}[c]{ CH1}
\psfrag{C2}[c]{ CH2}
\psfrag{C3}[c]{ CH3}
\psfrag{A}[c]{ $A(1,2)$}
\psfrag{a}[c]{ $ $}
\psfrag{C}[c]{ $C(3)$}
\psfrag{c}[c]{ $ $}
\psfrag{B}[c]{ $~B(2)$}
\psfrag{b}[c]{ $ $}
\psfrag{d}[c]{ $ $}
\psfrag{D}[c]{ $~~D(1,2)$}
\psfrag{e}[c]{ $ $}
\psfrag{E}[l]{ $E(1)$}
\scalefig{0.8}\epsfbox{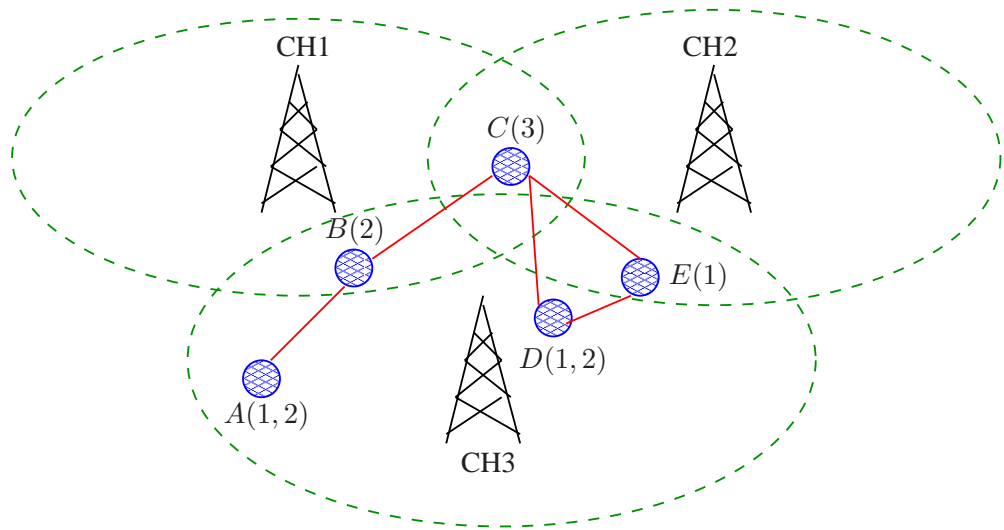}
\end{psfrags}}
\caption{Opportunity sharing among secondary users: graph coloring.}
\label{fig:OS}
\end{figure}

We illustrate the problem of spatial spectrum opportunity sharing with the help of
Figure~\ref{fig:OS}. Assume that there are three primary users, each occupying one of the three channels.
A secondary user within the coverage area of a primary user cannot use the channel occupied by that primary user.
For example, channels available to secondary user $A$ are $(1,2)$. Furthermore, neighboring secondary users
(indicated by a line connecting two secondary users in Figure~\ref{fig:OS}) interfere
with each other if they access the same channel. The problem is how
to allocate available channels to secondary users to optimize certain network utility such as sum capacity
under fairness constraints.

It has been shown in \cite{Zheng&Peng:05ICC,Wang&Liu:05VTC} that spatial opportunity allocation is equivalent to
graph coloring. Specifically, secondary users form vertices in a graph, and an edge between two vertices indicates
two interfering users. Treating each channel as a color, we arrive at a graph coloring problem: color each vertex using a number
of colors from its color list under the constraint that two vertices linked by an edge cannot share the same color. The objective
is to obtain a color assignment that maximizes a given utility function.

Obtaining the optimal coloring is known to be NP-hard
\cite{Garey&Johnson:book}. Centralized and distributed suboptimal approaches have been proposed \cite{Zheng&Peng:05ICC,Wang&Liu:05VTC}.
Game theory provides another approach to spatial opportunity allocation \cite{Halldorsson&etal:04}. An
interesting connection between the resultant colored graph and the Nash equilibria of the corresponding game
is noted in \cite{Halldorsson&etal:04}.

\section{Regulatory Policies}
\label{sec:policy}

\subsection{Aspects of OSA Policy}

Policy is obviously an important piece of OSA,
establishing rules of cooperation and joint usage between primary and secondary users.
In the US, the FCC is studying ways to advance secondary markets, such
as via interruptible leasing, a logical first step for commercial and user mutual
benefit.  A supporting policy could be fixed, or open to dynamic
negotiation and bidding; it could be centralized or decentralized.  Basic policy questions
such as these are impacted by a variety of factors, many noted above, depending on the application
and legacy systems.  It can be expected that intra-military systems, as well as
intra-commercial systems, can benefit
greatly from policies allowing spectrum sharing.  Should military and
commercial systems interact and coexist?  What form should such a policy take,
perhaps allowing for different modes of operation, such as in times of
national emergency?
Spectrum regulatory policies
vary over countries and regions, as well as across spectral sections.
How can policies be defined across international boundaries
and regions?  While it is generally
agreed that OSA can potentially bring numerous benefits, there are many technical,
as well as cost and business issues, to address before
widespread deployment can occur, and all these issues are intertwined with
policy.

Policies must be implemented on radio devices.
A logical argument for separation of the radio and the policy software includes
the option to add OSA capability into legacy systems, and the ability to update
or drop in new policies.  However, implementation of a separate software section
raises security and
software verifiability issues.  Modification by users could result in policy violations
\cite{Chapin-CommMag06}.  Further, device testing and verification for policy compliance
will be greatly complicated by dynamic policies and the complex interaction of many
devices sensing and reacting to the environment.

A wide range of policies are easily envisioned, spanning non-agressive to agressive,
or restrictive to permissive.  An obvious extreme is a do-no-harm policy, e.g.,
maintain complete orthogonality between systems at all times.  Less restrictive policies
may allow limited harm.  On the other extreme, in times of national emergency say,
a secondary system might have complete freedom to operate without restriction in an
otherwise occupied band.  In addition, while individual policies should be unambiguous,
it is easy to envision that multiple conflicting policies may arise.

An early example of a sense-and-respond OSA policy is DFS \cite{DFS-ETSI-05,DFS-ETSI-05b,FCC-03}.
DFS allows unlicensed
$802.11$ communications devices in the 5 GHz band to coexist with legacy radar systems.  The policy specifies
the sensor detection threshold, as well as timelines for radar sensing, usage, abandoning the channel, and a
non-occupancy time after detection.  This policy allows limited but minimal harm to
legacy radar systems, by accounting for the specific form of sensor for detection, and
prescribing the timelines for channel use and departure.  Another early example has been
developed in the DARPA NeXt Generation (XG) program, as a general listen-before-talk strategy,
analyzed by Leu et al. \cite{Leu-JNetMan06}.

\subsection{Where Policy Meets Signal Processing and Networking}

A policy, and compliance with a policy, is a function of specific parameters
available in a node.  What should be sensed, and what additional parameters
should be fed to the policy software to determine policy compliance?
Parameters may be raw or processed sensor outputs, environmental parameters,
and might include security codes or keys.  Environmental parameters might
include node identity, node location (e.g., from a GPS sensor), time
of day, as well as the location of nearby broadcasters, e.g., television.
Node location can be heavily leveraged, and policy could be both time and
location dependent, e.g., perhaps it would be desirable that elevated nodes
have more restrictive transmission
power levels during the day.  Location might also be used along with a
propagation model to estimate signal levels and their compliance with policy
such as a prescribed spectral level or mask.
The most fundamental sensor parameters
come from a power spectral estimate, providing a means to estimate spectral
occupancy.  Interesting extensions include the number and/or locations of other
nearby nodes, locations of cooperating nodes, types of message traffic,
priorities, delay constraints, and
observations about the environment other than just the power spectrum.
This is obviously not an exhaustive list of possibilities.
In addition to the sensing and environmental parameters, the radio may also
have a proposed action, whose compliance with policy may require verification.
For example, the frequency and transmit power of a possible transmission might
be suggested for permission to transmit.  Spectral masks may be employed to
determine acceptable levels of power both in-band and in adjacent bands due
to spectral leakage, as discussed in Section~\ref{sec:OFDM}.

The OSA process requires sensing, and a simple digital or analog
radiometer is a natural starting place (see Section~\ref{sec:detection}).  Some have coined the phrase
``interference temperature'' to refer to spectrum power levels and related
masks.  Along with temperature is the clear implication of a continuously
variable
level of interference. A power spectral density (PSD) estimator is easily implemented
and not too costly, and its statistics are well known.
Regulatory bodies, such as the FCC, will seek simple solutions based on
mature technology, that do not drive up device or systems costs, and are
easily understood.  However, even with this simplest of detectors, there are
a variety of subtleties in the sensing and interaction with policy.  While the
size of a frequency bin might be obvious in a given legacy case, wireless
brings the usual variety of complicating factors such as fading, widely varying local
propagation environments, hidden nodes, and indoor/outdoor applications.
There are typically many nonstationarities present in the wireless world,
raising such questions as what is the desired sensing (look-through) rate to
accommodate mobility, and what is the appropriate averaging time?  As described above, sensing is
inherently a probabilistic process, with the implication for harm in the OSA setting given
an incorrect hypothesis test outcome.
It is therefore interesting to contemplate policies that include probabilities,
or confidence levels.

As discussed in Section~\ref{sec:detection}, more sensitive and sophisticated detectors are available, such as those based on
cyclic statistics.  These are particularly effective for cyclostationary signals
such as digital broadcast, with long duration emission, and fixed known signal parameters.

The emergence of both sensor networks and MIMO technologies brings a variety of possible extensions
to the generic single sensor case.  A single node may incorporate array processing,
and thus spatial detection.  This will facilitate significant MAC improvements in
wireless networks, and can also facilitate OSA for array-equipped nodes.
In analogy to sensor networks, nodes may cooperate to perform distributed detection,
with many potential benefits.  Distributed detection approaches may overcome adverse
local fading effects to a large extent, but this approach adds communications overhead.
Multi-sensor techniques appear to complicate the policy definition, e.g.,
consider that the location and
number of cooperating nodes may be highly variable.
In addition, going beyond detection, signal parameters or features may be estimated, and signal
classifiers employed, such as modulation classification \cite{Swami&Sadler:00COM}.

Another approach to facilitating wireless networking is to take advantage of the
broadcast medium by employing beacons or control channels.  Beacons facilitate
medium access, and so facilitate channel sensing, and thus may ultimately prove
to be integral to the success of OSA schemes.
Beacons can be deployed to define permit-use or deny-use areas, and control channels
are integral to centralized systems such as cellular.  Thus beacons may be a simple
adjunct to facilitate OSA with legacy systems.   A drawback is that,
for military systems at least, the use of beacons is problematic from a security
and vulnerability
standpoint.

Thus, the type and capability of the sensor can play an integral role in policy,
and suggests a device-based policy.  This could be in a hierarchy, e.g., a more sophisticated
detector would enable
a more aggressive policy, because the more capable detector would presumably lead to better
decisions.

\subsection{Policy Reasoning}

Given a set of numerical parameters, it is straightforward to determine policy
compliance.  However, it may be highly desirable to have a policy reasoner (PR).
The need for reasoning
arises when a request is not posed in a y/n answerable form.  What frequencies are
available at power $P_0$ in frequency band $B_0$?  What constraints must be met to
allow a certain transmission?  This provides options for the radio, and requires
an interaction between the ``radio'' and the PR.
Interaction between the sensor/radio and the PR are highly desirable, and so defining
and standardizing this interface and possible interactive behaviors is important.
Wilkins et al. \cite{Wilkins-SRI-IEEEWC-06} have defined a policy reasoning language
specifically for OSA, and an interface
with three possible responses: ``yes'', ``no'', and ``yes with constraints''.  In the last case
the PR provides additional constraints that the radio would need to satisfy or order to enable the
requested transmission.  Example constraints are transmit power limit, transmit duration limit,
and so on.   This opens the question as to what is a rich enough set of constraints,
and this is subject to the particular device capabilities, e.g., whether it supports power control
and over what range.


Now is the time to carefully explore the interaction of policy, signal processing, and networking,
to study systems tradeoffs (complexity versus benefit, etc.), and to provide the tools
to the systems designers and the governmental policy makers.  Many specifics will be
determined by the nature of legacy systems.  Technologies will evolve to enable
more and more sophisticated signal processing, at cheaper cost, so policies will also
need to evolve.  This clearly motivates the
need for a drop-in policy reasoner approach.  Extensions must accommodate security
aspects, and further incorporate networking and network constraints.

\section{Concluding Remarks}

Opportunistic spectrum access is still in its infancy. Many complex issues in technical, economical, as well as
regulatory aspects need to be addressed before its potential can be assessed and realized.
Research efforts in the engineering community are particularly important in providing technical data
for the crafting of spectrum regulatory policies.

In this article, we have provided an overview of major technical and regulatory issues in OSA. Given
the complexity of the topic and the diversity of existing technical approaches, our presentation is by
no means exhaustive. We hope that this article provides a glimpse of the technical and regulatory challenges
of OSA and serves as an initial documentation of exciting research activities in the signal processing, networking,
and regulatory communities.

\bibliographystyle{ieeetr}
{\tiny

}

\end{document}